# Electrodynamics Correlates Knock-on and Knock-off: Current is Spatially Uniform in Ion Channels

Version 3

Robert S. Eisenberg
May 14, 2020

Department of Applied Mathematics
Illinois Institute of Technology

Department of Physiology and Biophysics
Rush University Medical Center

Chicago IL
USA

# Abstract

The electrodynamics of charge guarantee strong correlations between the movements of ions on all time scales, even those of atomic scale thermal motion. Ions in channels have been imagined as hard balls in a macroscopic mechanical model, for a very long time. Hard balls interact by collisions in such models, randomly knocking each other on and off 'binding' sites in thermal motion. But ions have large charge, and the hard balls of classical models do not. Correlations are present whenever Maxwell's equations apply, so correlations are present in individual trajectories, not just averages. Indeed, in a series system like an idealized narrow channel, the correlation is perfect (within the accuracy of Maxwell's equations) because of conservation of total current (that includes Maxwell's displacement current, the ethereal $\varepsilon_0\,\partial\mathbf{E}/\partial t$).

**Ethereal current prevents spatial variation of total current in a series system.**

Using macroscopic mechanical models to understand modern simulations (on the atomic length and time scale) gives complex results. The simulations show severe stochastic behavior in time and space, nearly that of a Brownian stochastic process with particles moving (on the average, crudely speaking) at the velocity of sound, some $10^{-9}$ meters in $10^{-12}$ sec.

**The stochastic complexity of spatial thermal motion disappears for current** in a series system. Total current does not depend on location in a series system like a narrow ion channel on any time scale.

**Spatial variables are not needed to describe total current in a one dimensional channel on any time scale, according to the Maxwell equations.** Removing the spatial dependence of total current should dramatically simplify theories and simulations, one might imagine.

An important challenge is the construction of a mathematical representation of current flow that takes advantage of conservation of total current, and avoids unnecessary detail, and the complexity of Brownian motion in space. The task is to actually construct such a simplified mathematical representation of total current and to see if that allows the calculation of channel properties at different voltages, at different concentrations, with different modulators, with different mutations, and with different structures.

# Introduction

Ion channels are narrow systems that accommodate ions close together in a space with the diameter of a few water molecules. Correlations produced by electrodynamics are not considered explicitly, if at all, in the classical view of ion channels. These correlations arise from the strong forces between the charges of the ions. The correlations are made good (i.e., enforced) on all time scales by the Maxwell equations of electrodynamics that require exact and universal conservation of current eq. (7), see ref. [1-3], as proven by the mathematics of the identity eq. (6) acting on the Maxwell eq. (3), giving the result eq. (7). In fact, correlation between ions in a one dimensional series system is unity, within the accuracy of the equations of electrodynamics, and so correlations dominate in nearly one dimensional ion channels. The correlations produce classical knock-on and knock-off behavior in the flow of ions—but not in the total flow of current—that is usually attributed to collisions. (A collision is a sudden change of direction of a trajectory. In a one dimensional system a collision is a reversal of direction occuring on a time scale much shorter than that of overall movement.)

Ions in channels have been studied for a very long time in another tradition [4], with a macroscopic paradigm appropriate for hard uncharged balls. Ions in channels have been imagined to move as uncharged particles through narrow pores, knocking into each other as they move, much as billiard balls did in the macroscopic mechanical model of Hodgkin and Keynes, with little if any discussion of correlations introduced by the electric field, Fig. 8 of ref [4][1]. Hille inherited [5, 6] and popularized this view [7]. Hodgkin and Keynes' ideas of knock-on and knock-off remain the word picture—the paradigm—used to analyze modern simulations of molecular dynamics [8-12] although modern simulations are computed for charged ions on a scale some $10^{15}$ times faster and $10^{8}$ times smaller than Hodgkin and Keynes' simulation.

**Fig. 1**

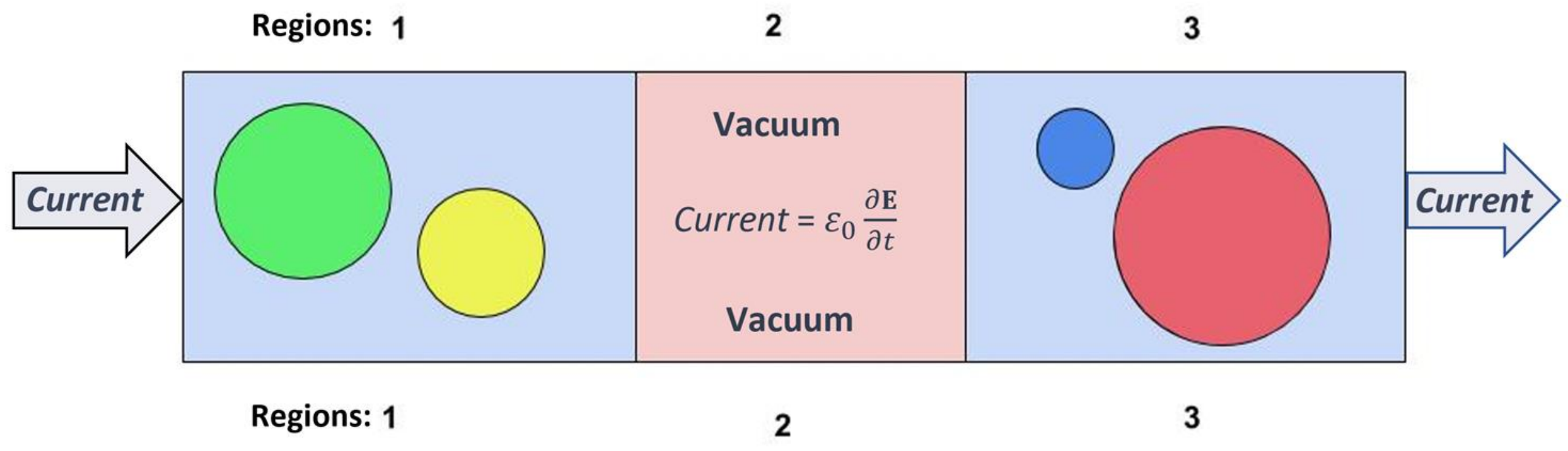

$Current = \mathbf{J}_{total}$ *in* **Regions 1, 2,** or **3**

Fig. 1**. Equal Currents** in a vacuum (Region **2**) and flanking solutions (regions **1** and **3**) in a one dimensional system. The central region labelled vacuum contains no matter, no atoms, nothing. Current through it is equal to the current in the end regions even though the current carriers are different in each region. Currents are equal on all scales including currents of individual atoms, even if Region **1** and Region **3** happened each to contain only one atom. Vacuum regions produced by dewetting play an important role in ion channels and are probably responsible for the sudden opening and closing of single channels, see references in text.

[1] analyzed on p. 81-84 with the help of A.F. Huxley.

The model shown in Fig. 1 includes a vacuum gap in Region **2**. Ion channels are thought to include vacuum gaps as documented below. Ions moving in Region **1** cannot move to Region **3** because of the vacuum gap. Nonetheless, ions moving in Region **1** produce current in Region **3.** The electric field produced by the charges in Region **1** spreads through the vacuum of Region **2** and moves the ions in Region **3.** The electric field spreads through the vacuum and links the motions of the atoms in Region **1** and the motion of atoms in Region **3**. **The atomic motions in Region 1 and Region 3 are perfectly correlated by the electric field, not by collisions of any type, not by knock-on knock-off behavior**.

Motions correlated by electric fields are not irregular in location at all, in a series system like Fig. 1. The total current flowing in this system is independent of location everywhere. The total current flowing in these regions varies dramatically in time because of thermal motion but it does not vary in space. **Maxwell's equations guarantee that the currents are equal at all times and on all scales in all regions of a series system** [1, 3], as we discuss and prove later in this paper, eq. (3)-(7). The time or ensemble averaged currents in each region are equal but so would be currents of individual atomic trajectories if Region **1** and Region **2** happened to contain only one ion each.

The components of the current are different in each region of Fig. 1, even though the total current is everywhere exactly the same, at all times. In regions **1** and **3,** current is produced by the movement of ions and also by the polarization (i.e., displacement) of charged matter, often drastically approximated by $(\varepsilon_r - 1)\varepsilon_0\,\partial\mathbf{E}/\partial t$.

In the vacuum region **2**, polarization occurs even though there is no matter and there are no atoms in region **2. This mysterious polarization of a vacuum is in fact a crucial property of the electromagnetic field**[2] that I think deserves the name 'ethereal'.

**Light is an ethereal current**. The polarization of the vacuum allows electromagnetic radiation like light to flow in an ethereal current from the sun to the earth. The polarization of the vacuum conducts an ethereal current $\varepsilon_0\,\partial\mathbf{E}/\partial t$ that is a property of space, not matter, and that exists everywhere, in a vacuum, in matter, even inside atoms, as described in textbooks of electrodynamics, electricity and magnetism, for example [13-16].

The ethereal current is part of the relativistic properties of space and time that make the charge on the electron independent of velocity, as velocities approach the speed of light, unlike mass, length, or time that change dramatically at high velocities. Many textbooks deal with the relationship of relativity and electrodynamics, including [17-22].

The ethereal current is **not** equal everywhere in the series circuit of Fig. 1. In the material Regions **1** and **3**, the ethereal current is likely to be a small fraction of the total current. In the vacuum Region **2**, the ethereal current is all the current. In the vacuum Region **2**, the ethereal current is equal to the total current in the material regions **1** and **3**, because the regions are all in series. None of the components of current are equal everywhere in the series circuit of Fig. 1. Only the total current is equal everywhere.

## Theory

**Variables**. In this paper, $\mathbf{J}_{total}$ is the entire source of the magnetic field $\mathbf{B}$ that appears on the right hand side of Maxwell's version of Ampere's law, our eq. (3). $\mathbf{J}$ is current carried by the movement of mass, however small, including what is usually crudely approximated as the material dielectric current $(\varepsilon_r - 1)\varepsilon_0\,\partial\mathbf{E}/\partial t$. Only the ethereal current $\varepsilon_0\,\partial\mathbf{E}/\partial t$ is isolated in

[2] and was recognized as such by Maxwell and his contemporaries.[24-26]

our field equations below. $\mathbf{J}$ includes all other currents. $\mathbf{E}$ is the electric field that measures the (mechanical) force on ions (for example). $\mu_0$ is the magnetic constant.

$\varepsilon_0$ is the electrical constant, the 'permittivity of free space'. $\varepsilon_r$ is the dielectric constant used to describe over-approximated dielectrics idealized as discussed in [23], [16], and classical texts. Ref [23] documents the difficulty with this representation of polarization by a single positive real number $\varepsilon_r$**.** Ref [1] shows several ways to write electrodynamics without this approximation.

$\mathbf{D}(x, y, z|t)$ is Maxwell's displacement vector field defined with the drastic approximation $\mathbf{D} = \varepsilon_r \varepsilon_0 \mathbf{E}$, where $\varepsilon_r$ is a single positive real number $\varepsilon_r \geq 1.0$.

The problematic nature of $\mathbf{D}$ and the polarization $\mathbf{P}$ is widely recognized. Nobel Laureate Richard Feynman calls their description of matter '*incorrect*' (on p. 10-7 of [27], my italics). Another Nobel Laureate Edward Purcell is even more critical saying *"the introduction of $\boldsymbol{D}$ is an artifice that is not, on the whole, very helpful. We have mentioned $\boldsymbol{D}$ because it is hallowed by tradition, beginning with Maxwell, and the student is sure to encounter it in other books, many of which treat it with more respect than it deserves."* (Purcell and Morin Ch. 10, p. 500 of [28], my italics.)

Purcell and Morin (on p. 507, my italics) go on to say even less charitably *"This example teaches us that in the real atomic world the distinction between bound charge and free charge is more or less arbitrary, and so, therefore, is the concept of polarization density* $\mathbf{P}$*."*

The same remark applies to $\mathbf{D}$ because $\mathbf{D}$ is defined as $\mathbf{D} \triangleq \varepsilon_0 \mathbf{E} + \mathbf{P}$. (1)

It should be clearly understood that **if the distinction between bound and free charge is more or less arbitrary, so is the entire treatment of the electric field, and electrodynamics, in essentially every textbook,** following [29-31].

It seems that the only way to remove this arbitrariness is to update the Maxwell equations so they do not depend on this definition of polarization at all, so they can be written in one form appropriate for any material system, no matter how its charge distribution varies with the electric field, or other forces. Ref. [1] attempts such an update.

**Special Properties of the Electric Field**. At first glance the field equations of the electric field seem similar to those of fluid dynamics. But they are not. Fluids do not exist in a vacuum; the electric and magnetic fields do.

In particular, magnetism has no obvious counterpart in fluid mechanics. Electricity cannot be separated from magnetism [32, 33], and so the field equations of electromagnetic dynamics are fundamentally different from the field equations of fluid dynamics. Feynman illustrates the relation of magnetism and electrostatics with simple but dramatic examples, ref. [27] on p. 13-8, paragraphs starting "Suppose … " and "Also …". These issues are discussed in Feynman (p. 13-6 ….) and less elegantly, later in this paper.

The fundamental source of the electric field is charge. Charge has markedly different properties from mass, its apparent analog in fluid mechanics. Charge is independent of velocity, even velocities approaching the speed of light. Charge is 'relativistically invariant'. Charge does not vary with velocity according to the Lorentz transformation of the theory of special relativity.[17-22] Mass varies with velocity according to the Lorentz transformation and so the equations describing the flow of mass are fundamentally different from those describing the flow of charge. Magnetism and light itself are consequences of this special property of charge, as is the ethereal current $\varepsilon_0\, \partial\mathbf{E}/\partial t$ described in textbooks of electrodynamics.

Magnetic fields $\mathbf{B}$ are generated by **total current**

$$\mathbf{J}_{total} = \mathbf{J} + \boldsymbol{\varepsilon}_0 \frac{\partial \mathbf{E}}{\partial t} \tag{2}$$

according to the following:

**Maxwell's version of Ampere's law**

$$\frac{1}{\mu_0} \mathbf{curl\ B} = \mathbf{J} + \varepsilon_0 \frac{\partial \mathbf{E}}{\partial t} \tag{3}$$

Many of the special properties of electrodynamics arise because the electric field changes shape as a solution of Maxwell's equations. A direct integration in time (with zero initial conditions) of eq. (3), gives the following, as discussed at length in [1, 3].

$$\mathbf{E}(x, y, z|t) = \frac{1}{\varepsilon_0} \int_0^t \left( \frac{1}{\mu_0} \mathbf{curl\ B} - \mathbf{J}(x, y, z|t') \right) dt' \tag{4}$$

**In one dimension, where curl $\mathbf{B} = \mathbf{0}$,**
the integration gives

$$\mathbf{E}(x|t) = -\frac{1}{\varepsilon_0} \int_0^t \mathbf{J}(x|t') dt' \tag{5}$$

*Please note that $\mathbf{J}(\cdot)$ is not $\mathbf{J}_{total}(\cdot)$*

The identity $\quad \mathbf{div\ curl\ B}(x, y, z|t) = \mathbf{0}$ (6)

implies the **Universal Exact Conservation of Total Current**

$$\mathbf{div\ J}_{total} = \mathbf{0} \tag{7}$$

Eq. (7) may look like an ordinary property of an incompressible fluid but it is not. Incompressible fluids are only incompressible in an approximation of say 0.01% over a reasonable dynamic range of forces. But the electric flow $\mathbf{J}_{total}$ being described by eq.(2), (3) & (7) has no measurable compressibility for voltages as small as say hundreds of nanovolts or as large as billions of volts. Both potentials are accessible in laboratories. The dynamic range of 'perfect' incompressibility is something larger than $10^{17}$.[3] The time scale of validity of eq.(3) & (7) is similarly beyond our normal experience, ranging from the shortest times we can measure to the time scale of light travelling to stars if not galaxies (depending on how the phenomenon of dark matter and dark energy are interpreted).

[3] Given measurements of astronomical forces in stars, and near galactic centers, the dynamic range is probably very much larger.

**Skepticism is appropriate**. The reader may agree with me that skepticism is appropriate as one thinks through the many extraordinary implications of eq. eq.(3) – (7). Skepticism has been expressed by almost all physicists confronted by the properties necessary to describe electromagnetic radiation flowing through a vacuum, i.e., eq. (3), going back to Maxwell himself [24-26]. That skepticism motivated this and my previous papers on this subject [1-3, 23, 34-37].

**Remedy for skepticism is mathematics**. The mathematics of the derivation leaves no choice in these matters, whatever the physical subtleties or social or psychological hang-ups of readers like myself. It is difficult to dispute that the mathematical identity eq. (6) acting on the Maxwell eq. (3), gives the result eq. (7).

More discussion of the derivations can be found in ref [1] near p. 28, eq. (27) and in (6) where the derivation uses the Bohm version of quantum mechanics. Of course, there may be inadvertent errors involved in these derivations, but the derivations are hardly novel. Similar derivations are in the literature for 150 years.

Other details are also in the literature. A long discussion of the microphysics of current flow in a series of different components—resistor, vacuum capacitor, semiconductor diode, vacuum diode, salt solution, wire—at different times is found near Fig. 2 of [34]. Various implications and different ways of allocating components of current and charge are found in [2, 35-37] and are briefly discussed later in this paper.

**Back to the channel** of Fig. 1: it is important to realize that the dominant form of interaction between ions are the forces described by eq. (4) & (5). Collisions are not the dominant form of interaction of ions in this system. Indeed, the ions in Regions 1 and 3 interact only through the electric field. Ions in these two regions cannot collide with each other.

The example involving a vacuum region may seem artificial when applied to actual physical or biological channels. It is not. Real ionic channels include regions of vacuum because of dewetting [38] and the filling and dewetting of these regions is thought to be responsible for (at least some kinds) of gating of single channels.[39-41] The sudden opening and closing [42, 43] of single channels, loosely called ‘gating’, is a central property—even defining feature—of all biological channels that are studied one channel at a time [44, 45]. The large literature of dewetting in channels can be accessed through ref [38].

In Fig. 1, the forces are not the same in each region or location. The forces on charges in regions **1** and **3** move the charges to make the total currents equal. The forces are defined by the electric field $\mathbf{E}(x)$ of eq. (5) in our one dimensional system where $\mathbf{curl\,B} = \mathbf{0}$. Maxwell’s equations in turn determine the $\mathbf{E}(x)$ that enforces the equality of total current.

The forces can be computed from the appropriate time dependent version of Coulomb’s law—a combination of the time dependent continuity equation and Maxwell’s first equation—if it includes all charges, of any kind whatsoever. The forces can also be computed directly as the solution eq. (4) of the Maxwell version of Ampere’s law, as explained in detail on p. 25 of ref [1]. In one dimension, eq. (5) is enough: $\mathbf{curl\,B}$ of the magnetic field is zero in a one dimensional system. In higher dimensions, the $\mathbf{curl\,B}$ field must be determined as well and that involves much more than the simple integration shown in eq. (5). The determination then requires the solution of the Maxwell equations and boundary conditions that describe the structure of the protein and how it evolves in time.

**One dimensional systems are remarkably simple**, as well as important. The simplicity of the one dimensional eq. (5) compared to the three dimensional eq. (4) is striking. In a series system like a channel or a diode, the one dimensional integral eq. (5) requires only knowledge of the total current. It requires knowledge of a single value—a function of time, but not a function of space. Current $\mathbf{J}_{total}$(*one dimensional*) can be evaluated at any convenient location because it is independent of location. That current is experimentally accessible. It can be measured by the voltage drop across a small resistor inserted anywhere in the series system. That current $\mathbf{J}_{total}$(*one dimensional*) can be determined by a theory, or numerical calculation, or simulation at just one location, including perhaps by a boundary condition that is simpler than field equations. On the other hand, using the three dimensional eq. (4) requires a complete solution of Maxwell's equation, in space and time, which is a formidable task in any situation, and more or less impossible in time varying protein structures, or realistic electronic circuits, for that matter.

A comparison of eq. (5) and eq. (4) suggests why so much of our technology and so much of biology can be described by one dimensional branched circuits. In one dimensional systems current (and energy) cannot 'leak into space'. Signals stay where they are supposed to be. Signals on one wire can be kept distinct from signals elsewhere. The $\mathbf{B}$ field that describes leakage into space does not have to be evaluated because $\mathbf{curl\,B}$ does not exist in one dimension.

Note that conventional circuit diagrams [46] do not show the return paths by which currents 'complete the circuit'. By convention they only show a perfect ground. In real circuits [47-49], the return paths for current force a certain component of two dimensional behavior. There is an area involved because the forward and return circuits are not in the same place. Indeed, branched one dimensional circuits also have analogous areas. These areas are small (in some sense) but they may generate significant $\mathbf{B}$ fields, in special cases, e.g., in ground planes and ground return circuits, and this may complicate and limit the use of these circuits, at least in my view, because $\mathbf{curl\,B}$ is no longer zero in two dimensions. Twisted pairs allow high speed data transmission and replaced coaxial cable, making the ethernet and internet possible at reasonable cost. The area between the conductors in the twisted pair is nearly zero and so the two dimensional properties are minimized as will be the **curl B** field, in my view.

**Molecular Dynamics.** Modern simulations of the dynamics of atoms [50-53] do not explicitly consider the conservation of current or the correlations it produces, as far as I know.[51, 54] They rather use Coulomb's law as their only description of their ***time varying dynamic*** electric field[4] despite Feynman's admonition, slightly paraphrased here) that "*Coulomb's law is false. It is **only to be used for statics.***" [(p.15-14 of [27]: my emphasis and italics.]

**Incomplete truths**. Indeed, Feynman goes on to use Coulomb's law as an example of the *"… danger in [using the scientific] process before we get to see the whole story. **The incomplete truths learned become ingrained and taken as the whole truth**. What is true and what is only sometimes true become confused."* [slight paraphrase; my emphasis and italics].

[4] Electric fields in molecular dynamics are not static. They vary on time scale typically represented as a discrete process with time steps of $\sim 10^{-15}$sec.

This statement might be the theme of this entire article and its predecessors [1-3, 23, 34-37]. They are devoted to removing the incomplete truths in the classical formulation of Maxwell's equations.

There can hardly be a more important example of "incomplete truths taken as the whole truth" than classical Maxwell equations, with their nearly arbitrary $\mathbf{D}$ and $\mathbf{P}$ fields. The confusion of "what is true and what is only sometimes true" is particularly evident in the misuse of the dielectric constant in nearly every textbook of electrodynamics.[23]

Modern molecular dynamics do further damage in their treatment of the electric field. Not only do they use only the static Coulomb's law to compute electrodynamics on a $10^{-15}$sec scale, they further approximate a general **non**-periodic electric field as a periodic system and use various conventions involving Ewald sums to evaluate the electric field. They usually compute the field with three dimensional Ewald sums derived classically from conditional infinite series, **without discussion of the Riemann rearrangement theorem and the lack of uniqueness of the value of any conditional infinite series**. The electric fields and transport are computed at equilibrium despite the presence of large atomic scale flows (evident in the trajectories of the simulation) and the presence of large macroscopic flows that are crucial to the natural function of the molecules being simulated. Numerical checks of the validity of the expressions used are notable by their absence, particularly compared to the numerical checks of simulations of particles in semiconductors, see Ref [55] Fig.6.34-6.36, p. 313-315, that do not use periodic boundary conditions or assumptions of zero flow. Indeed, some of that checking has been done for ions in water and published in detail [56-62].

The effects of these approximations on the (universal and exact) reality of conservation of current are not clear. Even the simplest quantities like the concentration (i.e., number density) of ions are affected. [The reader from the physical sciences needs to be reminded that the type and concentration of ions are dominant determinants of the function of many biological systems and molecules.] The periodic boundary conditions, for example, make it difficult even to define the concentration (i.e., number density) of ions precisely. Different research groups define concentrations with different conventions producing different results.[63] 'Conventions' are not part of mathematics or well-defined physical theories because they produce results that cannot be transferred from one research group to another or from one set of experimental conditions to another. They are called 'nontransferable' for that reason. Non-transferrable theories are not helpful in designing systems like integrated circuits that operate under a range of conditions.

**Currents reported in simulations** have great complexity in space and time, approximating the properties of a mathematical Brownian stochastic process that reverses direction an uncountably infinite number of times in any finite interval, no matter how brief. Maxwell's equations guarantee spatially uniform currents in a series system like a channel on all time scales, even for systems in thermal motion, involving a few atoms or just one, even inside atoms [3, 64-67], even for individual trajectories approximating a trajectory in a Brownian stochastic process.

Spatially uniform currents do not involve the location variable. They are much much less complex than currents that vary almost as Brownian processes in space. Indeed, a constant (in space) is about as different from a Brownian trajectory (in space) as it could be: a constant replaces a function of unbounded variation.

Here seems an opportunity to create a mathematics, or mathematical physics, that greatly simplifies the analysis of current flow in series systems, if this special property of total current can be exploited, as it has been in the semiconductor literature of one dimensional devices for a long time. Ferry [68, 69] provides a fine modern treatment of a classical literature [70] that includes [71-74]. This approach was extended to ions in channels by Schuss and collaborators [75] using the theory of stochastic processes (e.g., an adaption of renewal theory) to deal with the complicated shot noise produced when several types of ions move through channels, each with a different current voltage relation that is not independent. Each varies with concentration of every type of ion. Ethereal current $\varepsilon_0\, \partial\mathbf{E}/\partial t$ was not included in the work of Schuss and collaborators.

Branched networks are one dimensional systems of particular interest because they form the circuits of our digital electronics and computers. Kirchhoff's current law can be slightly generalized [2] to describe branched networks of one dimensional components. A further generalization could include the branched networks of our nerve cells and nervous system, and even much of the 'circuitry' of biochemical metabolism described in detail in classical biochemistry textbooks.

**Total current is simpler than ionic movement**. Current flow $\mathbf{J}_{total}$ is easier to understand and to compute, on all scales, than the motions of all atoms (as measured by flux $\mathbf{J}$). Ions move cooperatively, because of the electric field. **Collisions are not the dominant form of interaction of ions**. The dominant form of interaction is the perfectly correlated motion required so total current flow is conserved (exactly, everywhere on all time and length scales in which the Maxwell equations are valid).[1-3, 23, 34-37]

Models of biological function (that depend on total current [76]) can be easier to understand than models involving all atomic motions because of these cooperative movements, just as movements of current in the wiring of our home, or the transmission lines which connect our homes to a generator in a remote power plant, is much easier to understand than all the movements of charges involved in those currents.

**Correlations do not depend on averaging**. It is important to understand that these correlations occur in individual trajectories on all time and length scales whenever Maxwell's equations apply.

The correlations do not depend on averaging. In particular, the correlations are perfect in one dimensional series systems on atomic time scales.[5] They guarantee equality of current at all locations within ion channels, at all time scales, without temporal or spatial averaging, to the extent that ion channels are one dimensional systems. If the pore of the ion channel has regions without mass (e.g., vacuums without water or ions because of de-wetting [38], for example) current flow through that region is equal to total current flow everywhere else in the one dimensional system as described in previous paragraphs.

[5] The correlation function is in fact $1 - \delta$ in a one-dimensional series system, where $\delta$ is the accuracy of Maxwell's equations of electrodynamics.

These special properties of current in a one dimensional system arise from the electrodynamics of charge. The electrodynamics of charge are specified by partial differential equations that exist everywhere and specify forces everywhere, not just between particles. Electrodynamics is difficult to define using only particles because $\varepsilon_0\,\partial\mathbf{E}/\partial t$ is a property of (all) space, not just the space containing particles. Electrodynamics usually requires field equations, typically partial differential equations, to describe $\varepsilon_0\,\partial\mathbf{E}/\partial t$. Theories that include only the locations of charges have difficulty capturing $\varepsilon_0\,\partial\mathbf{E}/\partial t$ where the particles are not, i.e., almost everywhere.

In special cases, where structure changes very little, pre-charged capacitors—between acid and base side chains of proteins and the pore of the channel, for example—can be used to gain insight, using the divergence theorem, partial integration, and the mean value theorem to approximate the charge stored in the partial differential equations.[6] This is equivalent to using nonzero (time independent permanent) initial conditions on charge to determine the $t' = 0$ component (i.e., the $t' = 0$ lower 'limit' of the definite integral) in eq. (4) & (5).The pre-charged capacitors are an electrical engineer's way of avoiding the more awkward phrase "charge (per volt) in Coulomb's law".

**Updating Maxwell**. Textbooks hardly mention electrical currents produced by convection, diffusion, and (realistic) polarization [23] that are essential components of many biological (and technological) systems. Recent papers [1, 3] show one way to update the Maxwell equations to include these important currents while removing the nearly arbitrary definitions of polarization $\mathbf{P}$ and the displacement field **D.** Updating Maxwell forces a reassignment of the components of current. Ref. [1] tries to do this with minimal changes to the classical forms of electrodynamics found in textbooks for a long time [29-31, 77] so attention can be focused on the main issues of physics, and not the (somewhat) arbitrary definition of components. The definitions in Ref. [1] are certainly not unique, and not likely to be to the taste of all readers. But focus should be on the special properties of the electric field and the remarkable consequences of Ampere's law (Maxwell version eq. (3)), namely eq. (4), (5) & (7). These are independent of the choice of components.

**Current in a one dimensional series system**. The consequences of conservation of total current are particularly striking in a one-dimensional series system like an ion channel. There, conservation of total current means that **total current is equal everywhere, independent of location, at any time**.[7] Note that the flow of atoms $\mathbf{J}$ (strictly speaking the flux of mass that has charge) is **not** equal everywhere in a series circuit because mass and charge can accumulate when $\partial\mathbf{E}/\partial t \neq 0$.[35]

All who work with real circuits know the importance of $\partial\mathbf{E}/\partial t$. They know the importance of the dielectric charge that moves and accumulates in 'stray' capacitances according to $\varepsilon_r\varepsilon_0\,\partial\mathbf{E}/\partial t$ [2, 35], where $\varepsilon_r$ is the (dimensionless) dielectric constant. Note that $\varepsilon_0$ is an irreducible unavoidable component of $\varepsilon_r$. The dielectric constant $\varepsilon_r$ cannot be zero. It is larger than one $\varepsilon_r \geq 1$. The $\varepsilon_r = 1$ is degenerate and does not describe anything except a vacuum.

[6] Maxwell [24-26] used this approach (with the name 'specific inductive capacity') and Wolfgang Nonner showed it to me.

[7] To be more precise, the integral of current across the cross section of the channel is defined as the total current in a nearly one-dimensional system. The radial current is assumed to be independent of spatial location. It does not have to be assumed to be small, although it usually is small.

The effects of $\varepsilon_r\varepsilon_0\, \partial\mathbf{E}/\partial t$ are large in real circuits and cannot be missed. Indeed, if stray capacitances are not included in the analysis of circuits, the circuits do not perform as desired. The charge in the stray capacitance helps produce the familiar $1 - \exp(-\, t/RC)$ charging curve of the (prefect ideal) resistor $R$ to a step of current. $C$ is the capacitance associated with the polarization $(\varepsilon_r - 1)\varepsilon_0\, \partial\mathbf{E}/\partial t$, of an ideal dielectric plus the polarization accompanying the ethereal current $\varepsilon_0\, \partial\mathbf{E}/\partial t$. [35]

The ethereal current $\varepsilon_0\, \partial\mathbf{E}/\partial t$ is universal, unchangeable, and unavoidable. The customary characterization of $\varepsilon_r\varepsilon_0\, \partial\mathbf{E}/\partial t$ as a 'stray' property is rather misleading. The noun 'strays' is used in ordinary English to describe someone or something that wanders homelessly, without family, something changeable to be avoided. 'Stray capacitances' on the other hand, contain a component that cannot stray. They contain a component $\varepsilon_0\, \partial\mathbf{E}/\partial t$ that is unicersal, unavoidable and unchangeable. All resistors (and resistances) are associated with a capacitance produced by $\varepsilon_0\, \partial\mathbf{E}/\partial t$. Ref. [2, 35] show how to evaluate that ethereal capacitance and include it in electrical circuits.

**The ethereal current acts as a perfect low pass spatial filter, filtering spatial components completely.** The ethereal current prevents the knock-on knock-off behavior of total current.

Evaluating separately the two components of total current—namely the ethereal displacement current $\varepsilon_0\, \partial\mathbf{E}(\text{channel}; x)/\partial t$ and $\mathbf{J}(\text{channel}; x)$ that sum to make the total current $\mathbf{J}_{total}$, see eq. (2)—requires computations of all the forces between all the atoms in Fig. 1 for the times of interest (as described by eq. (5) and in detail on p. 25 of Ref. [1]).

The computation of the components $\varepsilon_0\, \partial\mathbf{E}(\text{channel}; x)/\partial t$ and $\mathbf{J}(\text{channel}; x)$ is much more difficult than the computation of the total current itself $\mathbf{J}_{total}(\text{channel})$, particularly given the:

(a) nearly eleven orders of magnitude between the time scale of atomic motion and most biological function

(b) enormous range and strength of the electric field.

(c) complexity of engineering structures and additional time dependence of complicated biological structures.

$\mathbf{J}_{total}(\text{channel})$ is easy to evaluate. $\mathbf{J}_{total}(\text{channel})$ can be evaluated or measured at any **single** convenient location because it is independent of location in a one dimensional series system like a channel, including at the boundary. Indeed, in theories and simulations a boundary condition often specifies the current at the boundary location, making its evaluation trivial. In experiments, it is usually possible to embed a low value resistor in series with the system and measure the electrical potential across the resistor.

**Gating Current of voltage dependent channels**. Calculations of the components of gating current (of the voltage sensor of a voltage activated channel) illustrate this point.[78] Calculations of gating current in full atomic detail [79] can use Langevin equations to avoid equilibrium assumptions in its molecular dynamics.

Computations of current flow in classical molecular dynamics are particularly difficult to understand, beyond the admonitions of Feynman presented previously on p. 6. Classical molecular dynamics assumes periodic boundary conditions even though **no current or flux can flow between boundaries at the same potential.** Classical molecular dynamics assumes equilibrium but no current or flux flows at equilibrium. So how can classical molecular dynamics calculate a gating current, or any current for that matter, without two kinds of inconsistency? Inconsistent calculations can be done different ways with different results, making comparison of results difficult between different research groups, and thus significantly hindering progress in understanding. In such situations, the science may not be a social process that converges to a single successful result, useful in understanding, designing, and constructing new systems.

Independent of these issues, molecular dynamics may be able to describe atomic motions correlated by electrodynamics if it were modified to accurately compute electrodynamic forces. If molecular dynamics were to accurately compute electrodynamic forces, and if the system being simulated has polarization that everywhere can be described by a single real dielectric constant (as described in detail in ref [1], eq. 17, for example), it can describe atomic motion, once the calculation is properly extended to the biological time scale.[8] But this calculation cannot be done with classical methods of molecular dynamics [50-53] used in hundreds or thousands of laboratories today.

If dielectric properties vary significantly with location or anything else, calculation of polarization forces requires calculations of many 'bodies', and thousands of terms are needed for each body in the usual multipole expansions [80], when charged bodies are close together. Scientists are sometimes surprised at just how slow convergence is of a multipole expansion when charges are near each other.

What is needed to provide atomic resolution of particle motion (in my opinion) is a molecular dynamics that exploits the special properties of total current, perhaps using the averaging-in-time methods introduced by Ma and Liu [81, 82] that yield the Poisson Nernst Planck model as an approximation [83]. Or perhaps by introducing a quasi-particle, a conducton that automatically satisfies equality of total current in a series system.

**Current in biology.** Current has been the main property of interest in many biological systems since the work of Hodgkin and Huxley [76], if not Hodgkin 1937 [84]. Current remains the focus of much biological interest [44, 45]: most of the channels in the nervous system and many in skeletal, cardiac, and smooth muscle seem to exist to pass currents, not to create fluxes of particular ions. These systems are not without general interest. Channel disfunction and the resulting abnormal conduction in cardiac muscle kills hundreds of thousands of people in the United States every year. Skeletal muscle makes up a large fraction of the mass of animals. The control of biological functions rests in the nervous system and in the channels that control its nerve cells and their connections.

[8] In implementing these calculations it may be useful to remember that potentials at short times are not well screened, because the ions of the system take some nanoseconds to respond. Dielectric screening is much faster but also much less effective. Crudely stated, potentials at short time reach to infinity while those at long times do not. Charges are so well shielded (by the ionic atmosphere at steady state) that ionic systems at long times behave as if they are electroneutral in some respects.[85,86]

The PNP class of models deal with current and electrodynamical correlations automatically, in the time dependent form of PNP, because the equations themselves satisfy the Maxwell equations. PNP is a nickname for the Positive Negative Positive 'doping' of a bipolar PNP transistor that might be analogous to a Positive Negative Positive spatial distribution of side change charge in a biological channel. PNP is also a nickname for the Poisson Nernst Planck equations. The pun—the double meaning of these nicknames—was intentional; indeed it was the main reason I chose this name.[87]

PNP models have played a crucial role in semiconductors (even with the limitations of its classical form) by describing the motion of its 'ions' (really, the quasiparticles called holes and 'electrons' [55, 88-93]) and is also important in physical and electrochemistry [94], despite its treatment of ions as points [95, 96]. PNP has recently been applied [97-104] to ion channels, allowing insight into the correlations determined by the electrical field.[99]

The classical PNP models need to be modified to deal with other types of correlations, for example, with the atomic scale correlations produced by steric exclusion, for example. Spatial correlations are included in many modern versions of PNP, for example, PNPF[9] [105-108], now renamed PNPB [109]. B stands for Bikerman whose elegant paper [110] seems to have started this approach.

PNPB is one of the more successful versions in dealing with bulk solutions, according to experimentalists [111]. PNPB has also been applied with some success to ion channels [105-108], including

(1) a model of the EEEE calcium channel used in many previous analyses;
(2) a model using the full structure of KcsA (Protein Data Bank PDB id **3F5W**); and
(3) a model using the full structure of the transporter NCX (Protein Data Bank PDB id **3V5U**).

The models are reduced. The atoms of the protein are do not move.

It should be clearly understood **that no matter what are the spatial correlations, the total current $\mathbf{J}_{total}$ will not vary with spatial location** in a series system like an ion channel. The ethereal current which is a solution of Maxwell's equations guarantees spatial uniformity of total current in a series system no matter what the other particle interactions, steric, knock-on or knock-off, or whatever, including chemical interactions involving redistribution of electrons in their orbitals around atoms in molecules.

[9] F stands for the Fermi-like distribution that accounts for correlations induced by spheres filling a channel or space.

# Discussion

**<u>Molecular Dynamics</u>**. It is not clear to what extent classical molecular dynamics [50-53] deals with the spatial correlations implied by conservation of current because of the complex codes used to approximate electrostatics and because of the inherent contradiction between current flow and periodic boundary conditions on electrical and electrochemical potential: obviously current does not flow between the boundaries of a computational cell which have the same potential, as they do in periodic systems.

More subtly periodic systems have been used, exploiting a toroidal (i.e., doughnut) geometry to allow current flow, but they involve artifacts as well: fluxes, not just current, are equal in these systems. Indeed, periodic boundary conditions impose exact equality on all co-ordinates of phase space not found in actual solids, on all time scales, even at a temperature of absolute zero, as pointed out sometime ago [112] and earlier in other publications unknown to me, no doubt.

In particular, periodic boundary conditions imply that location, velocity, acceleration, and cross coupling of all of those coordinates are periodic. Cross coupling is a difficult problem particularly in periodic approximations to systems with significant polarization current.

Polarization implies that every particle interacts with every other one. The number of such interactions is substantially—actually astronomically—larger than the number of particles in the system, particularly because pair-wise interactions are only a small fraction of the total number of significant interactions between polarizable particles. Indeed, it is not clear that a precise meaning can be given to the word 'state' in a system with polarization. If defining a state requires as many coordinates as defining the system itself, the state is nearly as hard to define as the system itself. Difficulties are not made easier by the uncountable infinity of locations needed to characterize the spatial properties of polarization. The idea of state is central to much of statistical physics and chemistry; the ethereal displacement current term $\varepsilon_0\,\partial\mathbf{E}/\partial t$ guarantees the universal existence of polarization current even in systems that do not include material polarization, for example systems that do not include $(\varepsilon_r - 1)\varepsilon_0\,\partial\mathbf{E}/\partial t$. These difficulties seem important in the reconciliation of electrodynamics and statistical mechanics, at least in its classical formulation.

**<u>Where are the checks and tests?</u>** In looking at the literature of molecular dynamics, what is striking is the dearth of direct tests that answer the obvious questions that have to be answered in any mathematically sound treatment of charged particles and their motions:

1) Do the simulations (of molecules dynamics of charged particles) satisfy the Poisson equation relating charge and electric field, or do they not?
2) Do the simulations (of molecular dynamics of charged particles) agree with analytical solutions (or numerical solutions of simple geometries known to be nearly exact) of the Poisson equation or not?

Comparison with known solutions of Maxwell's first and second equations are obviously needed, in my opinion.

Comparisons of this sort are the main topic of the numerical analysis of differential equations described in a voluminous literature. Simulations of molecular dynamics are solutions of differential equations, but they seem not to have been checked as other solutions of differential equations need to be checked. Checks of this sort are done routinely in particle simulations of computational electronics that solve differential equations quite similar to those in molecular dynamics. Ref [55] Fig. 6.34-6.36: p. 313-315 serves as an example and a modern entry point to the thousands of papers that document the work of the community working on computational electronics. There seems no logical reason similar tests cannot be performed on systems. Ferry and Ramey [113, 114] show how to apply these methods to static systems with finite size charges. Indeed, some of that checking has been done and published in detail [56-62].

Other methods combining dynamics and electrostatics have been developed [115-121], including many I am unaware of, no doubt. These methods also need to be tested directly to see if they yield exact equality of current in a series system, at all times, locations, and under all conditions. The results need to be compared with analytical solutions (or numerical solutions of simple geometries known to be nearly exact) of Maxwell's first and second equations.

**Coupling of flows in biology**. When modern molecular dynamics [50-53] and simulations are extended to include electrodynamics, and its simplifying correlations of ionic movement, more general insights into the mechanism of protein function are likely to emerge, particularly in processes like oxidative phosphorylation and photosynthesis in which electron flow is significant. Electron flows produce the chemical energy (in the form of ATP) which power almost all of the chemical processes of life. The coupling of electron flow and the movement of charged solutes (i.e., ions) is the crucial step in both oxidative phosphorylation and photosynthesis. The coupling is mediated by transporter proteins in the membranes of chloroplasts and mitochondria and is usually considered only on the atomic scale (say $10^{-10}$ m) of chemical interactions. But the electron flow is linked to the flow of ions by conservation of current and ***this may produce significant coupling, depending on the macroscopic structure and setup, in addition to the atomic scale 'chemical' coupling.***

Note the coupling in transporters will depend on the macroscopic structure and experimental setup in which the transporters and membranes function and are studied. Native organelles are 'finite cells' in which all membrane currents sum to zero, because there is no significant variation of potential within their cytoplasm [122-125]. Lipid membranes in experimental chambers are studied in voltage clamp mode, in which there is no requirement that the sum of all currents across the membrane be zero. Coupling of transporters measured in an artificial experimental setup is unlikely to be the same as coupling in native organelles where the transporters function.

These issues and the biophysical implications are profound because the coupling is a main mechanism of energy production and transduction in all of life. Coupling is a fundamental mechanism in nearly all active transport systems, and these are among the most important mechanisms in cells, in both health and disease.

It is important that coupling by conservation of current not be confused with atomic scale chemical coupling, if the mechanism is to be understood and controlled. Otherwise, artifacts can corrupt understanding of the atomic mechanisms. In macroscopic systems, part of coupling is determined by physiologically irrelevant choices in the design of the setup. That part of coupling should not be attributed to atomic mechanism of the transporter itself.

**Coupling of fluxes is multiscale**. The central point is that coupling is not simply a 'chemical' atomistic process of a particular transport protein or complex of proteins. It also depends critically on the geometry of the organelle or structure within which it is found and on the setup chosen by the investigator to study the transporter. Coupling is inherently multiscale because it involves the flow of (total) current. Current is described by the Maxwell equations that link all scales, see eq.(2), (3) & (7).

Conservation of total current couples everything in addition to coupling that occurs on the atomic scale of chemical interactions, as mentioned many times in this paper. These issues are discussed in more detail on p. 13 of [34], p. 23 of arXiv version; p. 11 of [35]; p. 10 of [2].

**Mathematics**. The question is how to include the perfect spatial correlation implied by conservation of current in our mathematical models and simulations of a series system. How do we take advantage of a property of the system that has no dependence on location?

One way is to reformulate variational principles like EnVarA [126] to use the Maxwell equations, or conservation of total current, to determine the electric field, instead of Poisson's equation. Another way is to apply conservation of total current as a constraint in the variational approach.

**Quasiparticles in channels and solutions**. Still another way might involve quasiparticles in the tradition of solid state physics, where the quasiparticles of semiconductors (holes and 'electrons'), superconductors ('Cooper pairs'), polarons, phonons, and so on, have been so useful.

We could construct conductons that are quasi-particles of total current (including the ethereal displacement current $\varepsilon_0\, \partial\mathbf{E}/\partial t$ of course). They would be complemented by the left over properties captured by polarons and diffusons, to deal with nonclassical polarization and concentration dependence, respectively. The resulting permion might indeed show simple knock-on, knock-off phenomena, analogous to those seen by Hodgkin and Keynes.

Permions [127] have been suggested to represent quasi-particles for ions permeating in channels. Ref. [128] shows some of the difficulties involved because different permeating ions follow different paths through the channel, twisting in three dimensions in different ways [127, 129], and thus requiring two quasi-particles I suspect. Each permion would not satisfy conservation of total current and so their spatial variation might allow knock-on and knock-off behavior. Only total current is immune from spatial variation.

**The central question** is this: can we use the special property of equality of total current as a lever to pry open the entire channel system? Can we determine the main biological function of many channels, the current through the channel as it varies with voltage, concentration, and structure of the channel, without dealing with the complexities of a spatial Brownian process, without dealing with knock-on and knock-off of particles in binding sites, to cite just one example?

The task is to actually construct such a simplified mathematical representation of current flow in general and in channels that takes full advantage of conservation of total current, and avoids unnecessary detail, particularly in one dimensional or series systems.

**Simplicity is indeed the goal**, but it cannot be found in treatments that ignore the electric field, as it is ignored in classical treatments of knock on and knock off in ion channels, widely used today. **The electric field creates correlations that simplify real systems**.[10] Maxwell's equations smooth the unbounded variation of a Brownian motion in space to a constant, the ultimate simplification, providing the greatest possible low pass spatial filtering in one dimensional channels.

**Simplicity is found when we study (total) current flow, not ion movement.**

## Acknowledgement

It is a pleasure to thank Jinn Liang Liu for criticisms that significantly improved the paper. David Ferry kindly introduced me to the literature of computational electronics showing no spatial dependence in its one dimensional systems, as he has to so much else.

Ardyth Eisenberg suggested that the title start by naming the main actor in the drama of channel function.

[10] Consider the correlations between the two ends of Kelvin's transatlantic telegraph cable [130-132]. We note that these systems contained no amplifiers, for readers unfamiliar with this classical, but obsolete technology.